\documentclass[a4paper,11pt]{article}
\pdfoutput=1 
\usepackage{jcappub}

\usepackage{graphicx}
\usepackage{longtable}
\usepackage{float}
\usepackage{dcolumn}
\usepackage{bm}
\usepackage{appendix}
\usepackage{multirow}
\usepackage{color}
\usepackage[utf8]{inputenc}
\usepackage{footmisc}
\usepackage{txfonts}
\usepackage{xcolor}
\usepackage{graphicx}
\usepackage{bm}
\usepackage{ulem}
\usepackage{multirow, array, float, tabularx}
\usepackage{amsmath}
\usepackage[colorlinks=true,linkcolor=blue,citecolor=blue]{hyperref}

\newcommand{\be}{\begin{equation}}
\newcommand{\ee}{\end{equation}}
\newcommand{\bea}{\begin{eqnarray}}
\newcommand{\eea}{\end{eqnarray}}
\newcommand{\bes}{\begin{subequations}}
\newcommand{\ees}{\end{subequations}}

\newcommand{\bc}{\begin{center}}
\newcommand{\ec}{\end{center}}

\usepackage{wasysym}

\begin{document}

\title{Constraining non-minimally coupled $\beta$-exponential inflation with CMB data}

\author[a]{F. B. M. dos Santos}\emailAdd{felipe.santos.091@ufrn.edu.br}

\author[b]{S. Santos da Costa}
\emailAdd{simonycosta@on.br}

\author[a,c]{R. Silva}
\emailAdd{raimundosilva@fisica.ufrn.br}

\author[d,e]{M. Benetti,}
\emailAdd{micol.benetti@unina.it}

\author[b]{J. S. Alcaniz}
\emailAdd{alcaniz@on.br}

\affiliation[a]{Departamento de F\'{\i}sica, Universidade Federal do Rio Grande do Norte, Av. Sen. Salgado Filho, 59078-970, Natal - RN, Brasil}

\affiliation[b]{Departamento de Astronomia, Observat\'orio Nacional Nacional, Rua Gen. José Cristino, 20921-400, Rio de Janeiro - RJ, Brasil}
	
\affiliation[c]{Departamento de F\'{\i}sica, Universidade do Estado do Rio Grande do Norte, Rua Almino Afonso, 59610-210, Mossoró - RN, Brasil}

\affiliation[d]{Scuola Superiore Meridionale, Universit\`{a} di Napoli ``Federico II'', Largo San Marcellino 10, 80138 Napoli, Italy}
 
\affiliation[e]{Istituto Nazionale di Fisica Nucleare (INFN) Sezione
  di Napoli, Complesso Universitario di Monte Sant'Angelo,  Edificio G,
  Via Cinthia, I-80126, Napoli, Italy}

\abstract{
The $\beta$-exponential inflation is driven by a class of primordial potentials, derived in the framework of braneworld scenarios, that generalizes the well-known power law inflation. In this paper we update previous constraints on the minimal coupled $\beta$-exponential model~\cite{Santos:2017alg} and extend the results also deriving the equations for the non-minimal coupled scenario. The predictions of both models are tested in light of the latest temperature and polarization maps of the Cosmic Microwave Background and clustering data. We also compare the predictions of these models with the standard $\Lambda$CDM cosmology using the Deviance Information Criterion (DIC), and find that the observational data show a moderate preference for the non-minimally coupled $\beta$-exponential inflationary model.
}

\maketitle

\section{Introduction}\label{sec1}

The inflationary scenario has become the current paradigm of the cosmology of the early universe. In the simplest description, it is driven by a single dynamical scalar field, predicting a nearly scale-invariant spectrum. Such a mechanism has proven to be able to solve a number of fine-tuning problems of the Standard Cosmology, and also a theory for the origin of structure in the universe that explains the Cosmic Microwave Background (CMB) anisotropies and the large-scale distribution of galaxies~\cite{Starobinsky:1980te,Guth1981,Linde1982,Albrecht1982}. 
Currently, cosmological observations give very strong support to the inflationary paradigm, showing a preference for plateau over monomial potentials of the inflaton field $\phi$, but still agreeing with several classes of models (see \cite{Martin2014n,Martin:2013tda,Rodrigues:2021txa,SantosdaCosta:2020dyl,Benetti:2019kgw,SantosdaCosta:2017ctv,Piccirilli:2017mto,Benetti:2016jhf,Benetti:2016ycg,Benetti:2016tvm, Benetti:2013cja} for recent analyses for different classes of models). 
On the other hand, the possibility of a non-minimal coupling of the inflaton to gravity has been investigated since the first models of inflation \citep{Futamase:1987ua,Lucchin:1985ip,Kaiser:1994vs}. These models consider a coupling term $\frac{1}{2}\xi\phi^2 R$, where $\xi$ is the strength of the coupling and $R$ is the Ricci scalar, which arises from quantum corrections in curved space-time, and is an ingredient for the renormalization of the scalar field in this context \citep{Faraoni:1998qx,Freedman:1974ze,Kaiser:2015usz}. Some examples are non-minimally coupled chaotic inflationary models with arbitrary potential in the framework of supergravity~\cite{Kallosh:2010ug} and  models driven by a non-minimally coupled Higgs field \cite{Bezrukov2008,Barvinsky:2008ia}, in a way that the strength of the coupling comes from the Standard Model itself  (see also \cite{Linde2011}). In the observational perspective, it has also been shown that in some cases the introduction of a non-minimal coupling may improve the description of the data (see e.g. \cite{Linde2011,Fakir1990,Komatsu1998,Komatsu1999,Campista:2017ovq,Ferreira:2018nav,Reyimuaji:2020goi,Bostan:2018evz,Tenkanen2017} and references therein).

In this paper we investigate minimally and non-minimally coupled inflationary scenarios in which the accelerated expansion of the early Universe is driven by a class of potentials that generalizes the well-known power law inflation through a general exponential function~\cite{Alcaniz:2006nu}. As discussed in \cite{Santos:2017alg}, the $\beta$-exponential potential can be derived from the framework of braneworld scenarios and is able to provide a good description of the observational data.  Specifically, our goal here is threefold: first, we update the analysis of the minimally coupled $\beta$-exponential model reported in \cite{Santos:2017alg} using the latest   Planck (2018) data; then, we introduce and discuss some cosmological consequences of the non-minimally coupled $\beta$-exponential scenario; finally, we perform a statistical analysis of the current CMB and clustering data to test the models predictions and use the deviance information criterion to compare the observational viability of the $\beta$-exponential scenarios with respect to the standard $\Lambda$CDM model. 

We organize this paper as follows. In Section \ref{sec2}, a brief review of theories with a non-minimally coupled scalar field is presented. The non-minimal coupled $\beta$-exponential model is introduced in Sec. \ref{sec3}, where we also present a slow-roll analysis considering both the minimal and non-minimal coupled scenarios. We discuss the observational data sets used in our statistical analysis and the method used in Sec. \ref{sec4}, while we present our results in Sec. \ref{sec5}. Finally, the main conclusions of the paper are summarized in Sec. \ref{sec6}.

\section{Non-Minimally Coupled Scalar Field}\label{sec2}

In this section, we review the treatment for a non-minimally coupled scalar field with gravity, showing the conditions for which slow-roll inflation happens in this context, as well as the main observables. In the Jordan Frame, the general action of a scalar field non-minimally coupled to gravity is

\begin{gather}
S=\int d^4x\sqrt{-g}\Big[ \frac{\Omega^2(\phi)}{2\kappa^2}R - \frac{1}{2}g^{\mu\nu}\partial_{\mu}\phi\partial_{\nu}\phi - V(\phi) \Big],
\label{1}
\end{gather}
where $\xi$ is the strength of the coupling with gravity, $R$ is the Ricci scalar, $\Omega^2\equiv 1+\kappa^2\xi \phi^2$, with $\kappa^2 = 8\pi G=M_p^{-2}$ ($M_p$ being the reduced Planck mass), and $V(\phi)$ is the inflationary potential.
Clearly the action is reduced to the minimally coupled one for $\xi=0$.
Due to the presence of $\Omega^2$, the equations of motion that emerge from this action are quite elaborated but we can rewrite them in a more familiar form by performing a conformal transformation in the metric. Considering the transformation
\begin{gather}
\hat{g}_{\mu\nu}=\Omega^2 g_{\mu\nu},
\label{2}
\end{gather}
the Ricci scalar can be redefined in terms of the new metric and therefore we end up with the action
\begin{gather}
S=\int d^4x\sqrt{-\hat{g}}\Bigg[ \frac{1}{2\kappa^2}\hat{R} - \frac{1}{2}F^2(\phi)\hat{g}^{\mu\nu}\partial_{\mu}\phi\partial_{\nu}\phi - \hat{V}(\phi)  \Bigg],
\label{3}
\end{gather}
with
\begin{gather}
F^2(\phi)\equiv\frac{1+\kappa^2\xi\phi^2(1+6\xi)}{(1+\kappa^2\xi\phi^2)^2},\label{4}\\
\hat{V}(\chi)\equiv\frac{V(\phi)}{(1+\kappa^2\xi\phi^2)^2}\label{5}.
\end{gather}
Redefining the scalar field as
\begin{gather}
\chi_{\phi}\equiv\frac{d\chi}{d\phi}=\sqrt{\frac{1+\kappa^2\xi\phi^2(1+6\xi)}{(1+\kappa^2\xi\phi^2)^2}},
\label{6}
\end{gather}
we finally obtain
\begin{gather}
S=\int d^4x\sqrt{-\hat{g}}\Big[  \frac{1}{2\kappa^2}\hat{R} - \frac{1}{2}\hat{g}^{\mu\nu}\partial_{\mu}\chi\partial_{\nu}\chi - \hat{V}(\chi) \Big]\;,
\label{7}
\end{gather}
which maps the action \eqref{1} in the Einstein frame. 

We assume a slow-roll regime, i.e. the inflaton field rolls slowly along the inflationary potential, and calculate the slow-roll parameters $\epsilon$, $\eta$ and $\zeta$, which we rewrite here for the potential Eq. \eqref{5}. The Einstein frame slow-roll parameters in terms of this potential are 

\begin{gather}
\epsilon = \frac{M_p^2}{2}\left( \frac{V_{\phi}}{V \chi_{\phi}}\right)^2, 
\quad \eta = M_p^2\left(\frac{V_{\phi\phi}}{V \chi_{\phi}^2} - \frac{V_{\phi}\chi_{\phi\phi}}{V\chi_{\phi}^3}\right),
\quad \zeta^2 = M_p^4\frac{V_\phi}{V^2\chi_\phi^2}\left( \frac{V_{\phi\phi\phi}}{\chi_\phi^2} -\frac{3V_{\phi\phi}\chi_{\phi\phi}}{\chi_\phi^3} + \frac{3V_{\phi}\chi_{\phi\phi}^2}{\chi_\phi^4}-\frac{V_\phi\chi_{\phi\phi\phi}}{\chi_\phi^3} \right).
\label{8}
\end{gather}
where both $\epsilon$ and $\eta$ should obey the conditions $\epsilon,|\eta|<<1$ when inflation takes place. Inflation ends when one of these conditions is violated e.g. $\epsilon(\phi_{end})=1$. 

The inflationary parameters that are usually confronted with data are the spectral index $n_s$, being the variation of the scalar power spectrum with scale at horizon crossing, the tensor-to-scalar ratio $r$, as the ratio between the tensor and scalar spectrum amplitudes, and the running of $n_s$, which is the variation of $n_s$ with scale, also at horizon crossing, given by $n_{run}$; they can be expressed in terms of the slow-roll parameters as~\cite{liddle,Lyth_1999}
\begin{gather}
n_{s}=1-6\epsilon + 2\eta, \quad r=16\epsilon \quad \mbox{and} \quad n_{run} = 16\epsilon\eta - 24\epsilon^2 - 2\zeta^2
\label{9}
\end{gather}
The power spectrum of scalar perturbations can be computed when the CMB modes cross the horizon at the pivot scale $k_{\star}$
\begin{gather}
P_{R\star}=\frac{V(\phi_{\star})}{24\pi^2\epsilon M_p^4}\Bigg|_{k=k_{\star}},
\label{10}
\end{gather}
with $\phi_{\star}$ being the value of the inflaton field at horizon crossing. We set the value of $P_{R}$ at $k_{\star} = 0.05$ Mpc$^{-1}$ as $P_{R}=2.0933\times 10^{-9}$. The Planck Collaboration (2018) \cite{Akrami:2018odb} also determined the spectral index as $n_s=0.9649\pm0.0042$ (68\% C.L.), while the tensor-to-scalar ratio is constrained by the Planck + BICEP2/Keck Array data to $r<0.056$.

\section{The $\beta$-exponential model: slow-roll analysis}\label{sec3}

In this section we introduce the non-minimally coupled $\beta$-exponential model and 
study its slow-roll predictions. In order to make a comparison between the minimally and non-minimally coupled models, we begin reviewing the former case.

\subsection{Minimally coupled model}

\begin{figure*}[t]
	\includegraphics[width=8cm]{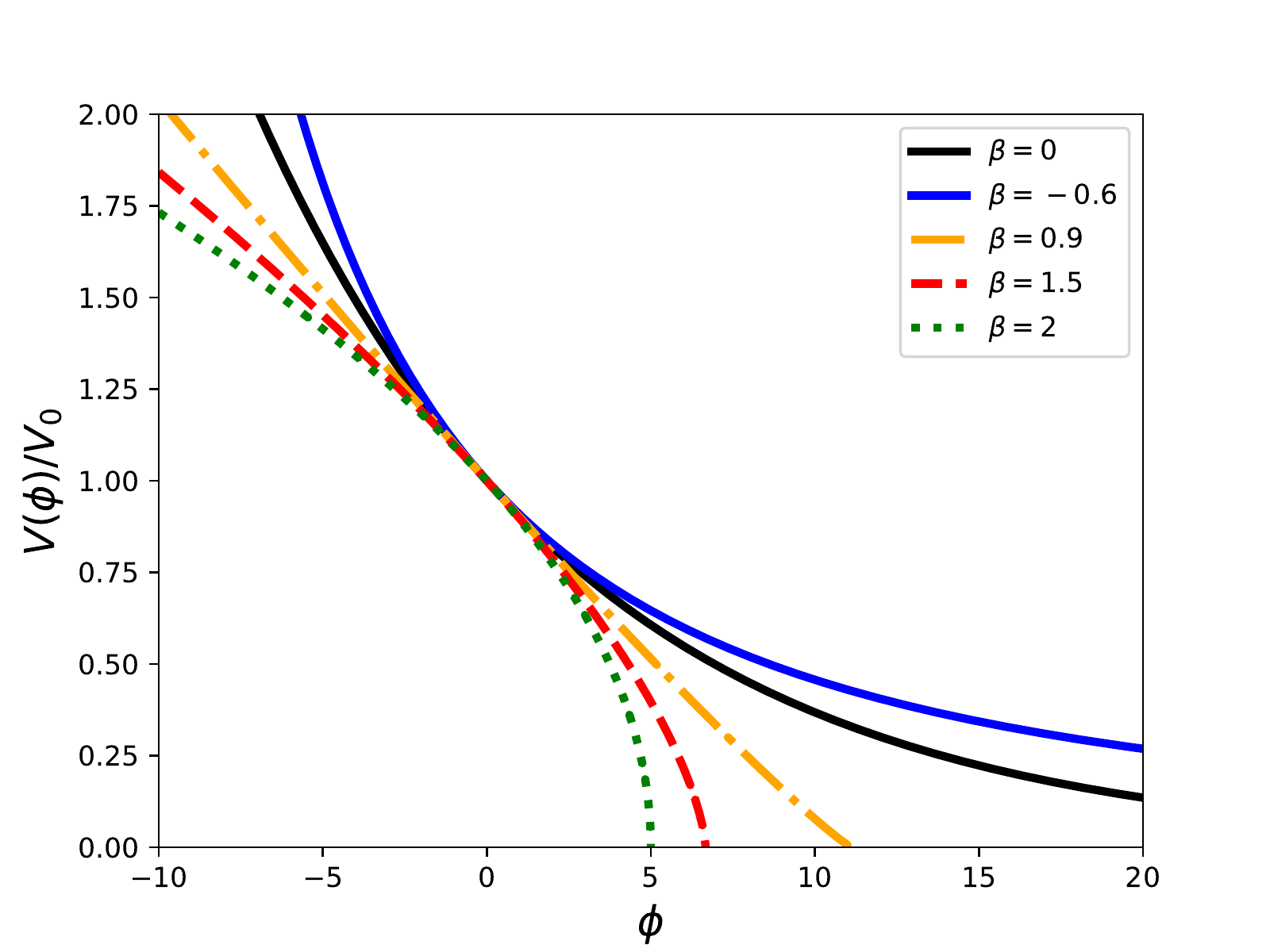}
	\includegraphics[width=8cm]{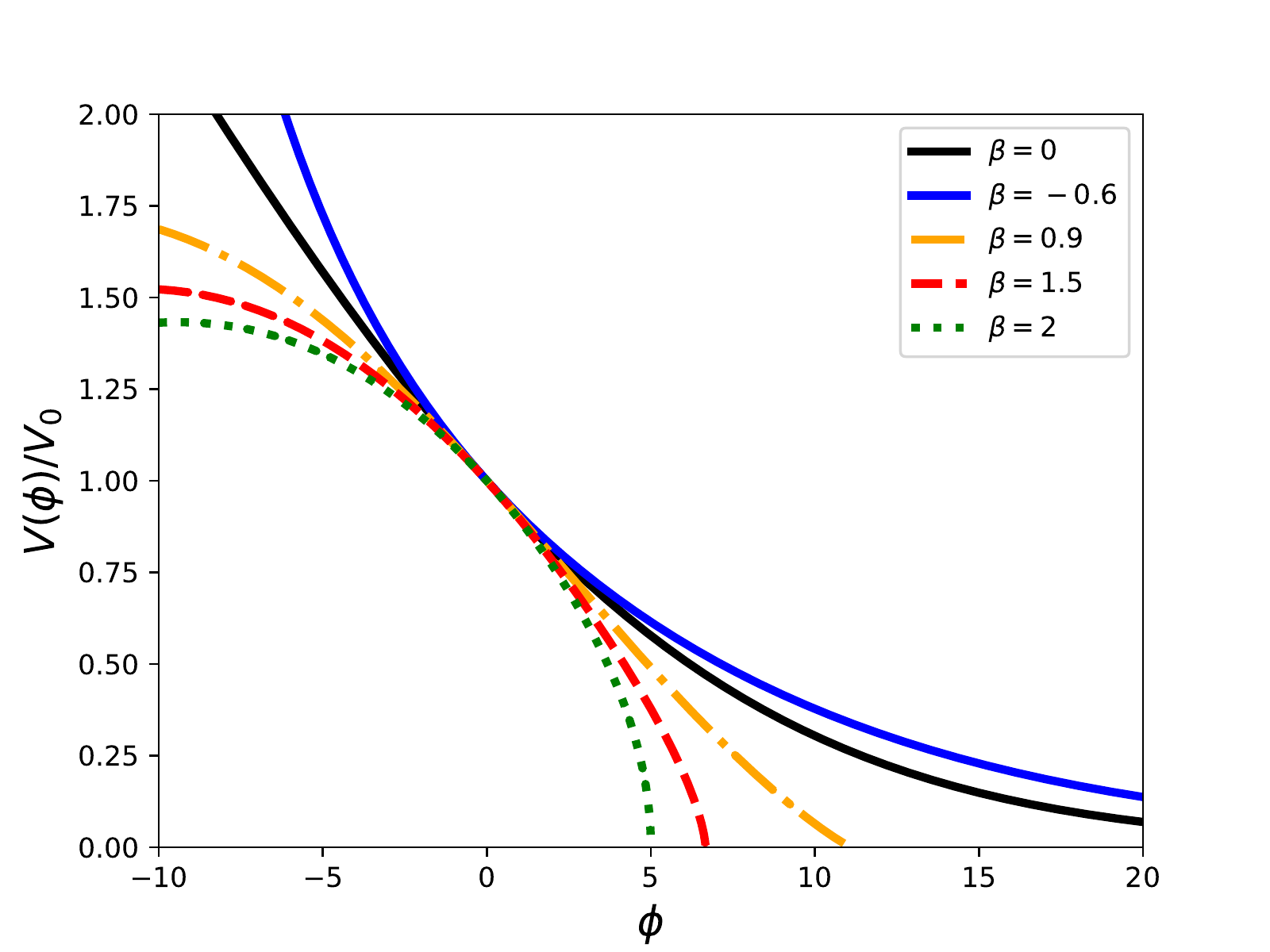}
	\caption{ The $\beta$-exponential potential. On the left, we show the minimally coupled function Eq. \eqref{12}, while on the right, we plot the non-minimally coupled function Eq. \eqref{17}, with $\xi=0.001$. The colors represent different values of $\beta$, and the value $\lambda=0.1$ was fixed in both panels.}\label{fig1}
\end{figure*}

The $\beta$-exponential potential
\begin{gather}
V(\phi)=V_{0}\left( 1 - \lambda\beta\frac{\phi}{M_p} \right)^{\frac{1}{\beta}},
\label{12}
\end{gather}
was introduced in \cite{Alcaniz:2006nu}, being subsequently motivated in the braneworld framework in  \cite{Santos:2017alg}. From the mathematical standpoint, the expression \eqref{12} is a generalization of the exponential potential through the parameter $\beta$, where in the limit $\beta\rightarrow 0$ we recover the usual exponential function. This function is defined such that when $1-\lambda\beta\phi/M_p>0$, $V>0$, and if $1-\lambda\beta\phi/M_p\leq0$, $V=0$. The behaviour of the potential \eqref{12} is shown in the left panel of Fig. \eqref{fig1}, where we can clearly see how the parameter $\beta$ controls the deviation from the exponential function. Note also that for $\beta=1/2$, one reproduces the quadratic chaotic inflation model, with the minimum shifted by $\phi=1/\lambda\beta$.

\begin{figure*}[t]
    \centering
	\includegraphics[width=8cm]{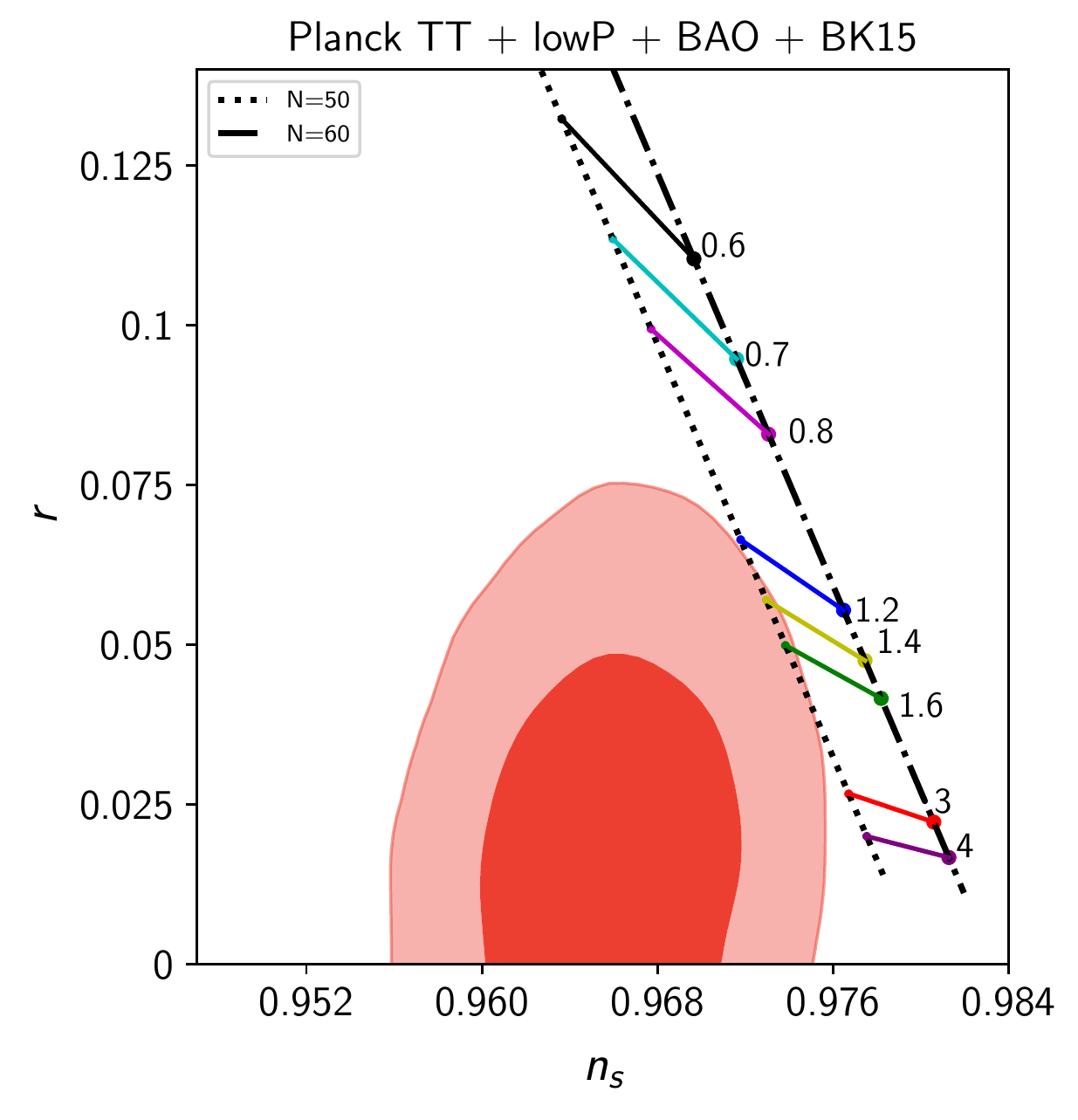}
	\caption{ The $n_s-r$ plane  for  different values of $\beta$ satisfying  Eq.~\eqref{16} and considering two values for the number of e-folds, $N= 50$ and $N= 60$.  The contours correspond to the Planck (2018)+BAO+BICEP2/Keck data ($68\%$ and $95\%$ C.L.) using the pivot $k_\star$= 0.05 Mpc$^{-1}$. }\label{fig2}
\end{figure*}

Setting the Planck mass to unity, the slow-roll parameters are given by 
\begin{gather}
\epsilon = \frac{\lambda^2}{2(1-\lambda\beta\phi)^2}, \quad \eta = \frac{\lambda^2(1-\beta)}{(1-\lambda\beta\phi)^2}\;,
\label{13}
\end{gather}
and we can find the value of the field at the end of the slow phase, $\phi_{end}$, by imposing $\epsilon=1$ and reversing the first of the above equations
\begin{gather}
\phi_{end} = \frac{1}{\beta}\Bigg[\frac{1}{\lambda}-\frac{1}{\sqrt{2}}\Bigg]\;.
\label{14}
\end{gather}
At the same time, by solving the integral for the number of e-folds $N_{\star}=\int_{\phi_{end}}^{\phi_{\star}}\frac{d\phi}{\sqrt{2\epsilon}}$, we find that the field value at horizon crossing $\phi_{\star}$ is
\begin{gather}
\phi_{\star} = \frac{1}{\beta\lambda}-\frac{1}{\beta}\sqrt{1/2+2\beta N_{\star}}\;,
\label{15}
\end{gather}
while the spectral index and tensor-to-scalar ratio can be written as
\begin{gather}
n_s -1 = -6\epsilon+2\eta =-\lambda^2\frac{(1+2\beta)}{(1-\beta\lambda\phi_{\star})^2}, \quad r = 16\epsilon = \frac{8\lambda^2}{(1-\beta\lambda\phi_{\star})^2}.
\label{16}
\end{gather}
By substituting Eq. \eqref{15} into \eqref{16}, we find that the predictions for both $n_s$ and $r$ show no dependence with $\lambda$, although this parameter will influence other quantities, such as the temperature power spectrum\footnote{In Ref.~\citep{Santos:2017alg}, it was shown the $n_s-r$ plane for different values of $\lambda$. However, the two plots displayed there are equal, as $n_s -1 = -2\frac{(1+2\beta)}{1+4\beta N_\star}$, and $r=\frac{16}{1+4\beta N_\star}$. We verified that this correction does not change the results reported in the paper \citep{Santos:2017alg}.}. This result is clearly shown in the $n_s - r$ plane displayed in Fig. \eqref{fig2}. 
From this analysis, we also find that using the current CMB temperature and polarization data from Planck (2018) combined with Baryonic Acoustic Oscillations measurements (BAO), only a small range of values of $\beta$ is marginally consistent at $95\%$ (C.L.), unlike the good agreement obtained from the Planck (2015) temperature and polarization likelihoods~\cite{Santos:2017alg}.
Indeed, it was previously found that $\beta=1.92\pm 0.05$ (68\% C.L.) for $\lambda$ fixed at $=0.07$, with a slight statistical preference for the $\beta-$exponential scenario over the $\Lambda$CDM model. In contrast, when $\lambda$ was left as a free parameter, the model was disfavoured with respect to the reference $\Lambda$CDM model.

\subsection{Non-minimally coupled model}

For the non-minimally coupled $\beta-$exponential model, the Einstein frame potential is given by
\begin{gather}
\hat{V}(\chi(\phi))=\frac{V_{0}( 1 - \lambda\beta\phi)^{1/\beta}}{(1+\xi\phi^2)^2}.
\label{17}
\end{gather}

The behavior of the non-minimal coupled potential can be seen in the right panel of Fig.\eqref{fig1}. We notice that for positive values of the field the form of both minimal and non-minimal coupled potentials is very similar whereas for negative values of $\phi$, a change in the shape of the curves is evident. This particular feature is characteristic of models with this type of coupling \cite{Tenkanen2017} and will play an essential role in the predictions of the model. The introduction of the coupling $\xi$ results in a change in the shape of the potential, as well as in the allowed  values of the $n_s$ and $r$ parameters.

From the slow-roll parameters in Eq. \eqref{8}, we obtain:
\begin{gather}
\epsilon = \frac{(4\beta\lambda\xi\phi^2 - \lambda\phi^2\xi - 4\phi\xi - \lambda)^2}{2(\beta\lambda\phi-1)^2(6\xi^2\phi^2 + \xi\phi^2 + 1)},\label{18}
\end{gather}
\begin{equation}
\begin{split}
\eta & = \frac{1}{(\lambda\beta\phi-1)^2(6\phi^2\xi^2+\phi^2\xi+1)^2}\Big[96\lambda^2\beta^2\phi^6\xi^4-48\lambda^2\beta\phi^6\xi^4  +6\lambda^2\phi^6\xi^4 -192\lambda\beta\phi^5\xi^4 +42\lambda\phi^5\xi^4 \\ & +96\phi^4\xi^4+16\lambda^2\beta^2\phi^6\xi^3-8\lambda^2\beta\phi^6\xi^3+\lambda^2\phi^6\xi^3  -32\lambda\beta\phi^5\xi^3+7\lambda\phi^5\xi^3-60\lambda^2\beta\phi^4\xi^3 +12\lambda^2\phi^4\xi^3 \\ & +16\phi^4\xi^3 +48L\phi^3\xi^3+12\lambda^2\beta^2\phi^4\xi^2-17\lambda^2\beta\phi^4\xi^2+3\lambda^2\phi^4\xi^2-24\lambda\beta\phi^3\xi^2 \\ & +14\lambda\phi^3\xi^2-12\lambda^2\beta\phi^2\xi^2+6\lambda^2\phi^2\xi^2+12\phi^2\xi^2+6\lambda\phi\xi^2-4\lambda^2\beta^2\phi^2\xi \\ & -10\lambda^2\beta\phi^2\xi+3\lambda^2\phi^2\xi+8\lambda\beta\phi*\xi+7\lambda\phi\xi-4\xi-\lambda^2\beta+\lambda^2\Big].
\label{19}
\end{split}
\end{equation}
Inflation ends when $\epsilon=1$ in such a way that we can recover the scalar field at the end of inflation by numerically solving \eqref{18} for $\phi_{end}$.

The expression for the number of $e$-folds at horizon crossing $N_{\star}$ is given by
\begin{gather}
N_{\star}=\int^{\phi_{\star}}_{\phi_{end}}d\phi\frac{\hat{V}}{\hat{V}_{\phi}}\chi_{\phi}^2.
\label{20}
\end{gather}
from which we can compute the field at the horizon crossing. For the potential of Eq. \eqref{17}, we find 
\begin{gather}
N_{\star} = -\int_{\phi_{end}}^{\phi_{\star}}d\phi \frac{\left(\lambda\beta\phi-1\right)\left(1+\xi\phi^2\left(1+6\xi\right)\right)}{\left(1+\xi\phi^2\right)\left(4\lambda\beta\xi\phi^2-\lambda\xi\phi^2-4\xi\phi-\lambda\right)}.
\label{21}
\end{gather}
Similarly to $\phi_{end}$, the scalar field at the horizon crossing, $\phi_{\star}$, should be obtained numerically for values of $\beta$, $\lambda$ and $\xi$ satisfying Eq.~\eqref{21} at the pivot scale $k=0.05$ Mpc$^{-1}$, i.e. $N_{\star}=55$. The amplitude of the scalar potential, $V_0$, is obtained by inverting Eq. \eqref{10}:

\begin{figure*}[t]
\centering
    \includegraphics[width=7.5cm]{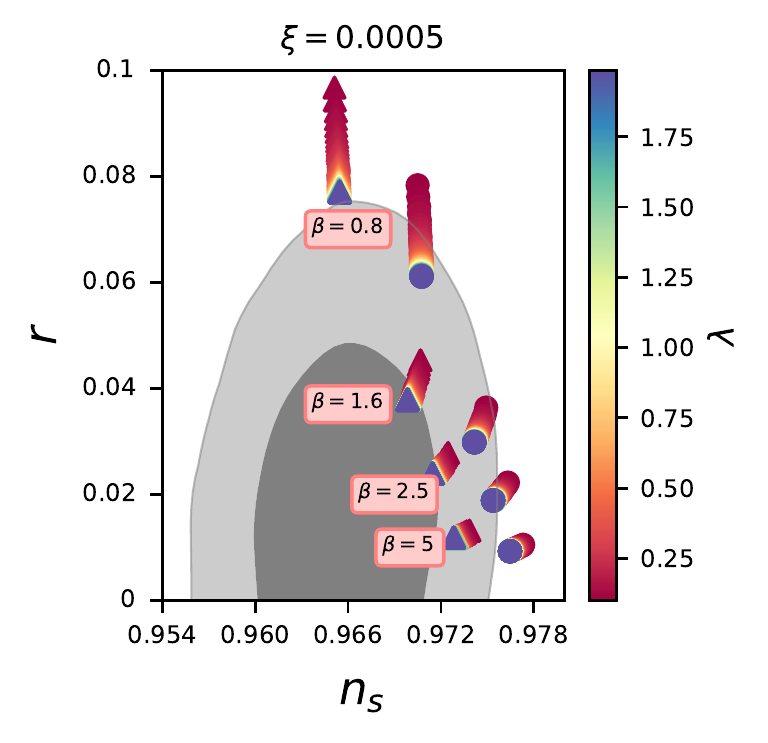}
    \includegraphics[width=7.5cm]{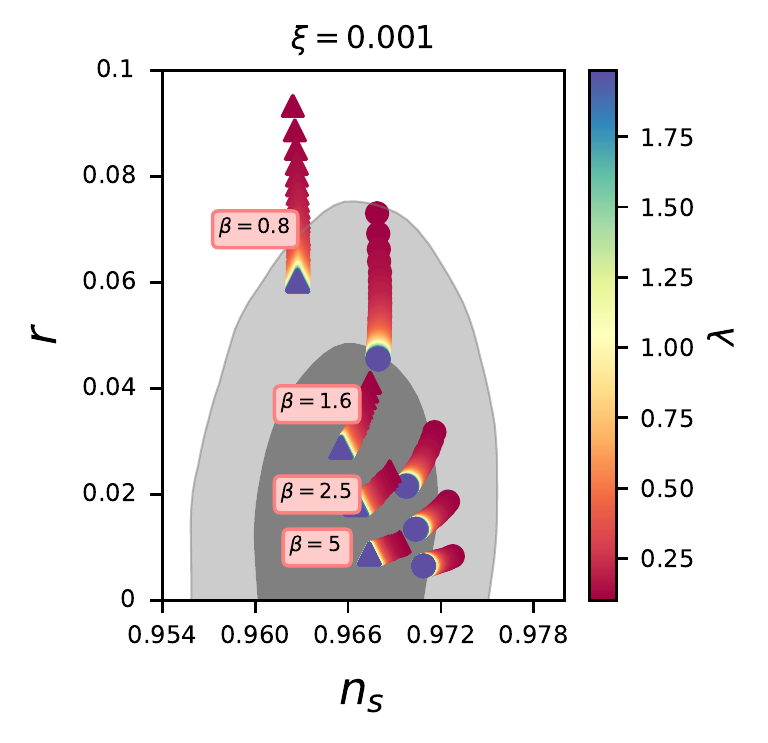}
    \includegraphics[width=7.5cm]{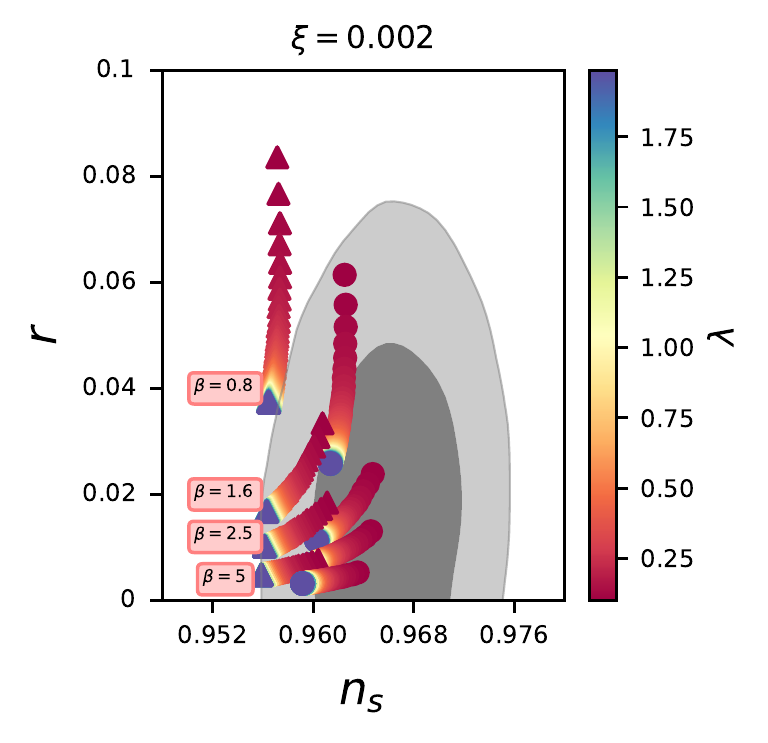}
    \includegraphics[width=7.5cm]{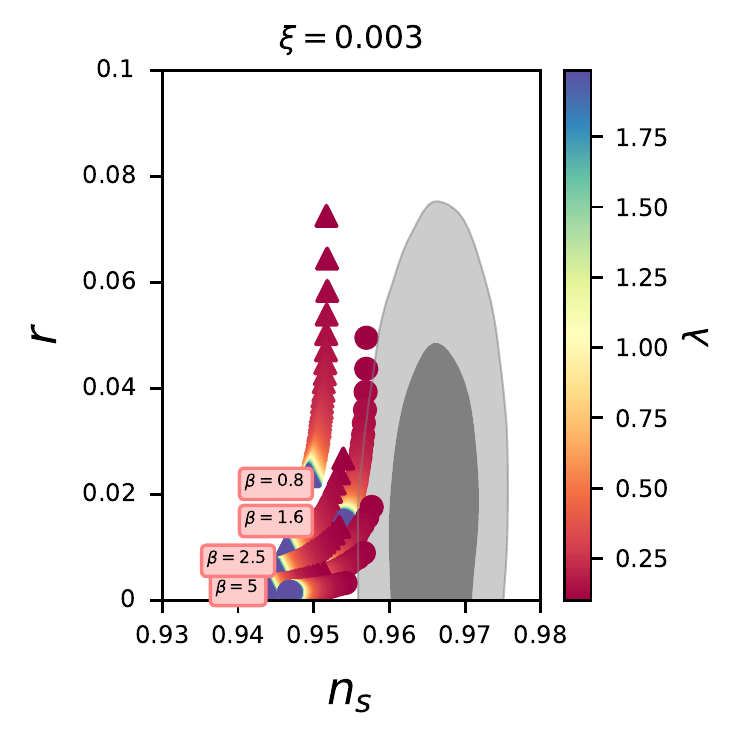}
    
	\caption{The $n_{s}-r$ plane for the non-minimally coupled $\beta$-exponential model. We consider four values of $\xi$, and the curves are expressed as a function of $\lambda$. In each plot, we also take different values of $\beta$, spanning the whole range of interest, as indicated in the labels. The triangular pattern corresponds to $N_\star=50$ predictions, while the circular one corresponds to $N_\star=60$ predictions.}
	\label{fig3}
\end{figure*}

\begin{gather}
V_{0} = \frac{1}{1 + \phi_{\star}^2 \xi (1 + 6 \xi)}\left\{ 12\pi^2 (1 - \beta \lambda \phi_{\star})^{-2 - \frac{1}{
		\beta}}  P_{R\star}(1 + \phi_{\star}^2 \xi)^2 \Big[\lambda +
\phi_{\star} (4 + \lambda (\phi_{\star} - 4 \beta \phi_{\star})) \xi\Big]^2 \right\}.
\label{22}
\end{gather}

Finally, the scalar spectral index and the tensor-to-scalar ratio can be written as 
\begin{gather}
r = 16\epsilon =\frac{8(4\beta\lambda\xi\phi_{\star}^2 - \lambda\phi_{\star}^2\xi - 4\phi_{\star}\xi - \lambda)^2}{(\beta\lambda\phi_{\star}-1)^2(6\xi^2\phi_{\star}^2 + \xi\phi_{\star}^2 + 1)},\label{23}
\end{gather}
\begin{equation}
\begin{split}
n_{s} & = 1 - 6\epsilon + 2\eta \\
& = 1 - \frac{1}{(\beta \lambda \phi_{\star}-1)^2 (1 + \phi_{\star}^2 \xi (1 + 6 \xi))^2}\Big[96 {{\lambda}^{2}} {{\beta}^{2}} {{\phi}_\star^{6}} {{\xi}^{4}}-48 {{\lambda}^{2}} \beta {{\phi}_\star^{6}} {{\xi}^{4}}+6 {{\lambda}^{2}} {{\phi}_\star^{6}} {{\xi}^{4}} -192 \lambda \beta {{\phi}_\star^{5}} {{\xi}^{4}}+60 \lambda {{\phi}_\star^{5}} {{\xi}^{4}} \\ & +96 {{\phi}_\star^{4}} {{\xi}^{4}}+16 {{\lambda}^{2}} {{\beta}^{2}} {{\phi}_\star^{6}} {{\xi}^{3}}-8 {{\lambda}^{2}} \beta {{\phi}_\star^{6}} {{\xi}^{3}}+{{\lambda}^{2}} {{\phi}_\star^{6}} {{\xi}^{3}}-32 \lambda \beta {{\phi}_\star^{5}} {{\xi}^{3}}   +10 \lambda {{\phi}_\star^{5}} {{\xi}^{3}} -24 {{\lambda}^{2}} \beta {{\phi}_\star^{4}} {{\xi}^{3}}+12 {{\lambda}^{2}} {{\phi}_\star^{4}} {{\xi}^{3}} \\ & +16 {{\phi}_\star^{4}} {{\xi}^{3}}+48 \lambda {{\phi}_\star^{3}} {{\xi}^{3}}+24 {{\lambda}^{2}} {{\beta}^{2}} {{\phi}_\star^{4}} {{\xi}^{2}}-14 {{\lambda}^{2}} \beta {{\phi}_\star^{4}} {{\xi}^{2}}+3 {{\lambda}^{2}} {{\phi}_\star^{4}} {{\xi}^{2}}  -48 \lambda \beta {{\phi}_\star^{3}} {{\xi}^{2}} +20 \lambda {{\phi}_\star^{3}} {{\xi}^{2}}  \\ & +24 {{\lambda}^{2}} \beta {{\phi}_\star^{2}} {{\xi}^{2}}+6 {{\lambda}^{2}} {{\phi}_\star^{2}} {{\xi}^{2}}+24 {{\phi}_\star^{2}} {{\xi}^{2}}-12 \lambda \phi_\star {{\xi}^{2}}+8 {{\lambda}^{2}} {{\beta}^{2}} {{\phi}_\star^{2}} \xi-4 {{\lambda}^{2}} \beta {{\phi}_\star^{2}} \xi+3 {{\lambda}^{2}} {{\phi}_\star^{2}} \xi  \\ & -16 \lambda \beta \phi_\star \xi+10 \lambda \phi_\star \xi+8 \xi+2 {{\lambda}^{2}} \beta+{{\lambda}^{2}}\Big],
\label{24}
\end{split}
\end{equation}
from which one may check that in the  limit $\xi=0$ the expressions of $n_s$ and $r$ for the minimally coupled case (Eq. \ref{16}) are recovered.

Theoretical predictions for $n_{s}$ and $r$ considering two numbers of e-folds, $N=50$ and $N=60$, are exhibited in Fig.~\eqref{fig3}. There we compare theoretical predictions for different values of $\beta$, $\lambda$, and $\xi$ with the $68\%$ and $95\%$ (C.L.) contours obtained from Planck(2018)+BAO+BICEP2/Keck Array data. Note that higher values of $\beta$ and $\lambda$ predict values of $n_s$ and $r$ in agreement with observations within $68\%$ (C.L.). On the other hand, higher values of $\xi$ shift the $n_s$ predictions and provide only marginal consistency within $68\%$ (C.L.), as seen in the $\xi=0.003$ case (lower right panel of Fig. \eqref{fig3}).

Furthermore, an effect of including a non-minimal coupling to the $\beta-$exponential potential is to make its predictions for the parameters $n_{s}$ and $r$ in better agreement with observations, which is clearly seen when we compare the Figs. \eqref{fig2} and \eqref{fig3}. The minimally coupled model is only marginally consistent with current observations for a small range of the values of $\beta$ (Fig.~\ref{fig2}) whereas the predictions of the non-minimal coupled model show good agreement with the observational data within $68\%$ C.L. (Fig.~\ref{fig3}).

\section{Method and analysis}\label{sec4}

We now introduce the method used to perform our analyses. First of all, we implement the theory described in the previous sections in the code {\sc ModeCode}~\citep{Mortonson:2010er,Easther:2011yq,Norena:2012rs} in order to obtain the theoretical predictions of the minimally and non-minimally coupled models for the CMB angular power spectrum. 
 
Given the form of the inflaton potential $V(\phi)$, the equations of inflationary dynamics, namely the Friedmann and Klein-Gordon equations, are integrated numerically in the code to obtain the Hubble parameter as function of the scale factor $a$, and the inflaton field as function of time. These quantities are also used to calculate the Fourier components associated with curvature perturbations produced by the fluctuations of the inflaton, which are solutions of the Mukhanov-Sasaki equations~\cite{Mukhanov:2005sc,Weinberg:2008zzc}
\begin{eqnarray}
u''_k+\left(k^2-\frac{z''}{z}\right)u_k=0\;,
\label{25}
\end{eqnarray}
where $u\equiv -z\mathcal{R}$ and $z\equiv a\dot{\phi}/H$, and $\mathcal{R}$ is the comoving curvature perturbations. Lastly, the primordial power spectrum of curvature perturbations $\mathcal{P}(k)$ is related to $u_k$ and $z$ via:
\begin{eqnarray}
\mathcal{P_R}(k)=\frac{k^3}{2\pi^2}\left|\frac{u_k}{z}\right| ^2.
\label{26}
\end{eqnarray}
Therefore, {\sc ModeCode} evaluates the spectrum of curvature perturbations when the mode crosses the horizon and then calculates the spectrum of CMB temperature fluctuations.

\begin{figure*}[t]
    \centering
	\includegraphics[width=7.5cm]{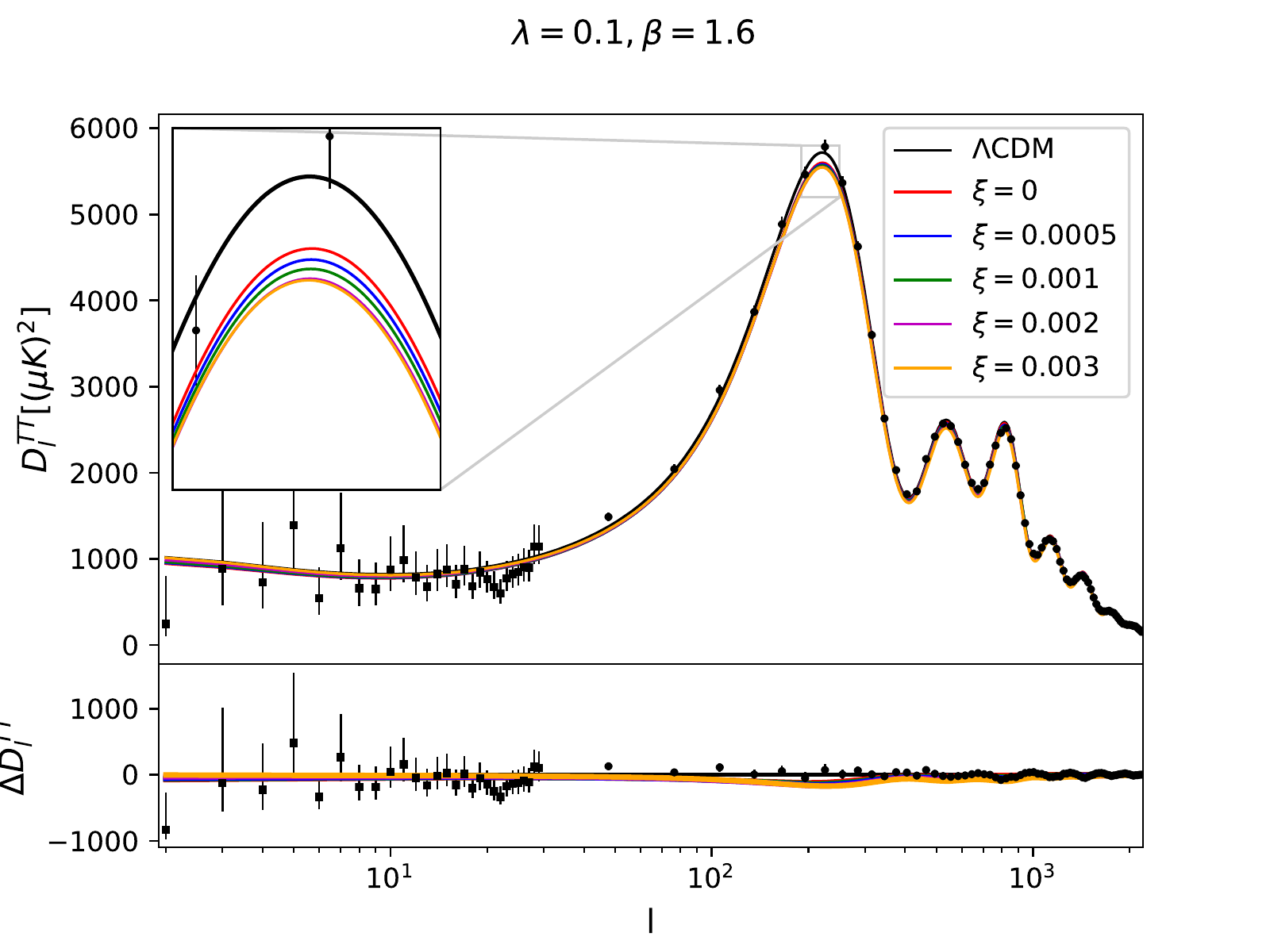}
	\includegraphics[width=7.5cm]{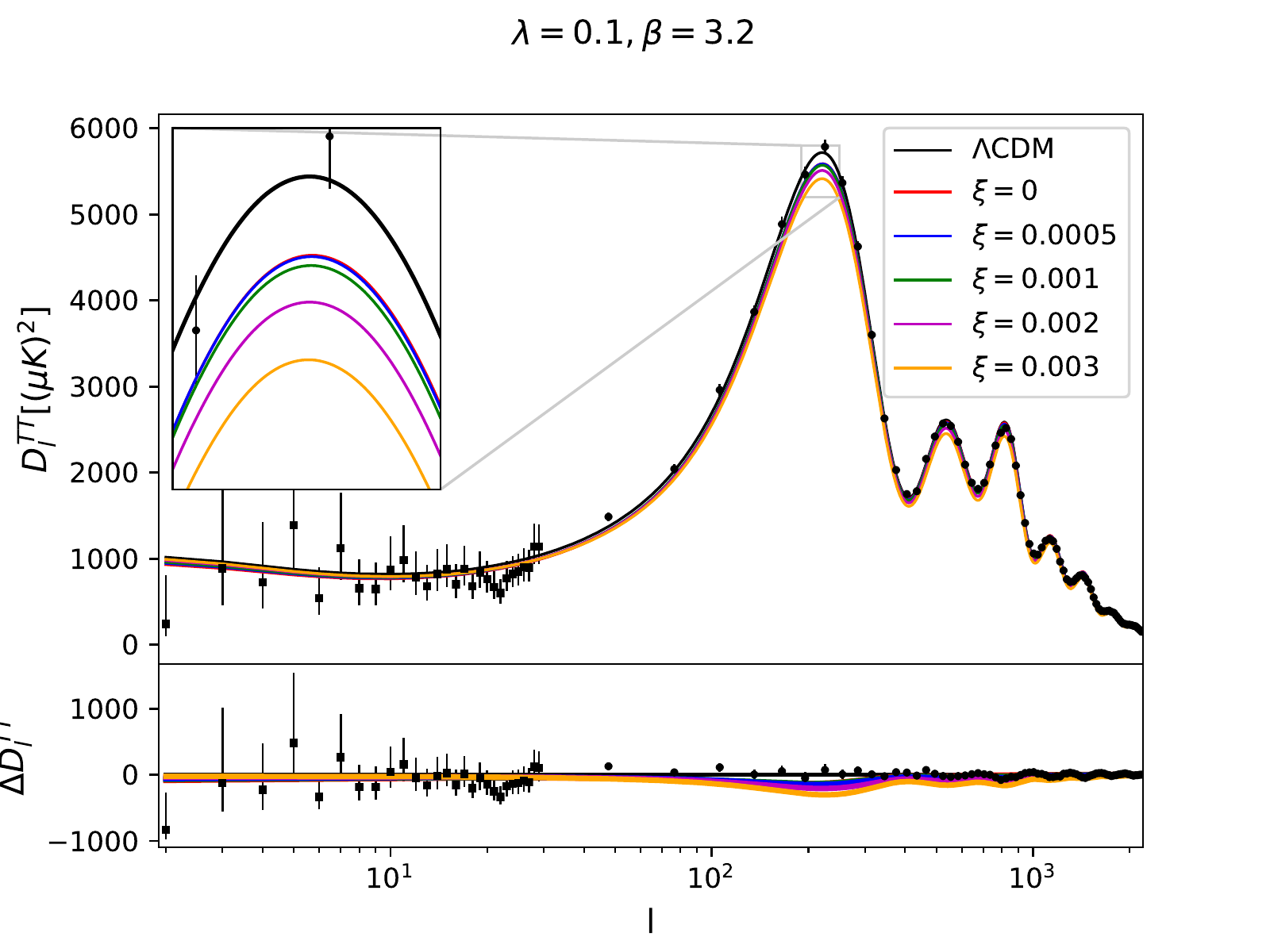}
	\includegraphics[width=7.5cm]{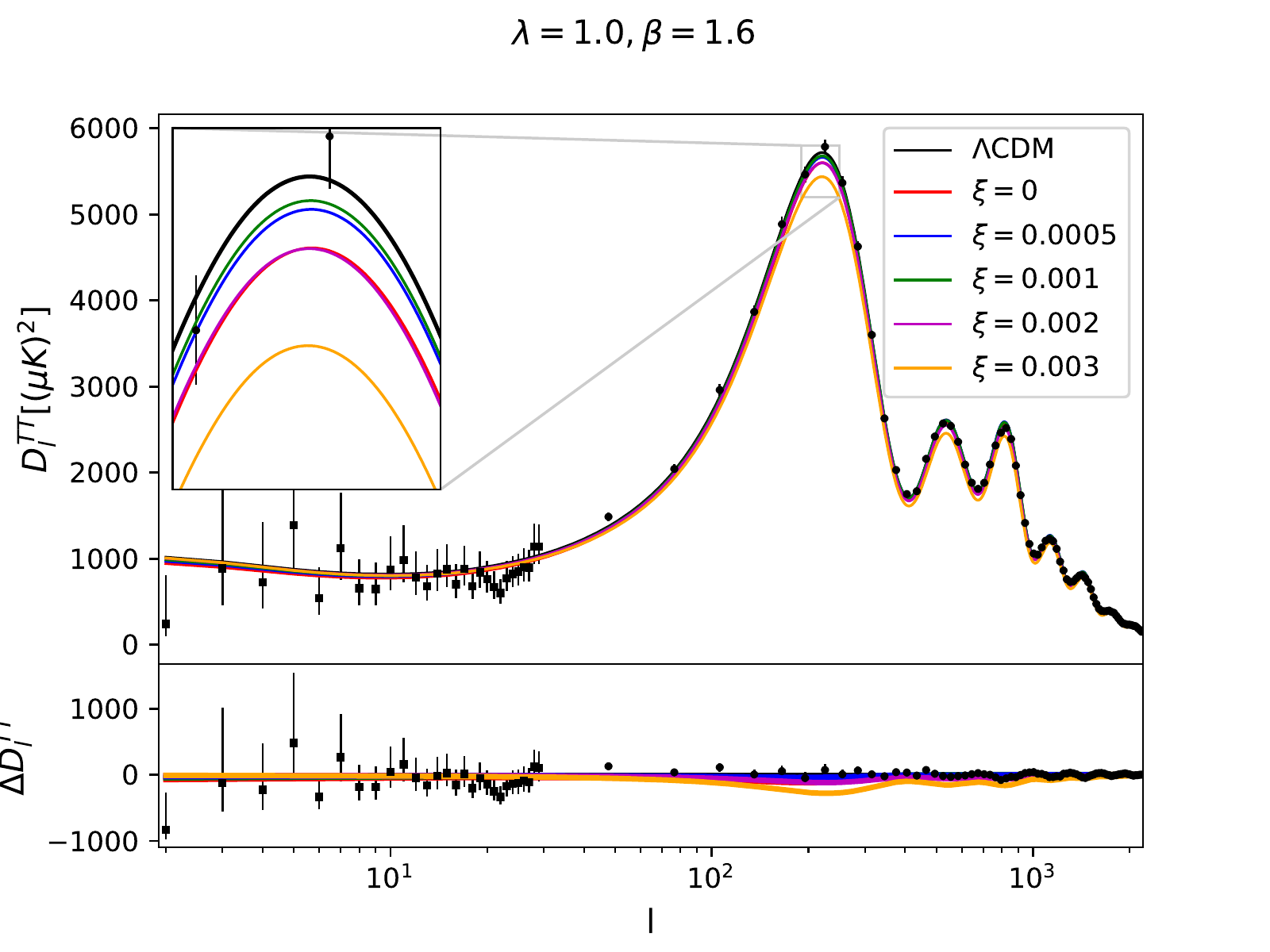}
	\includegraphics[width=7.5cm]{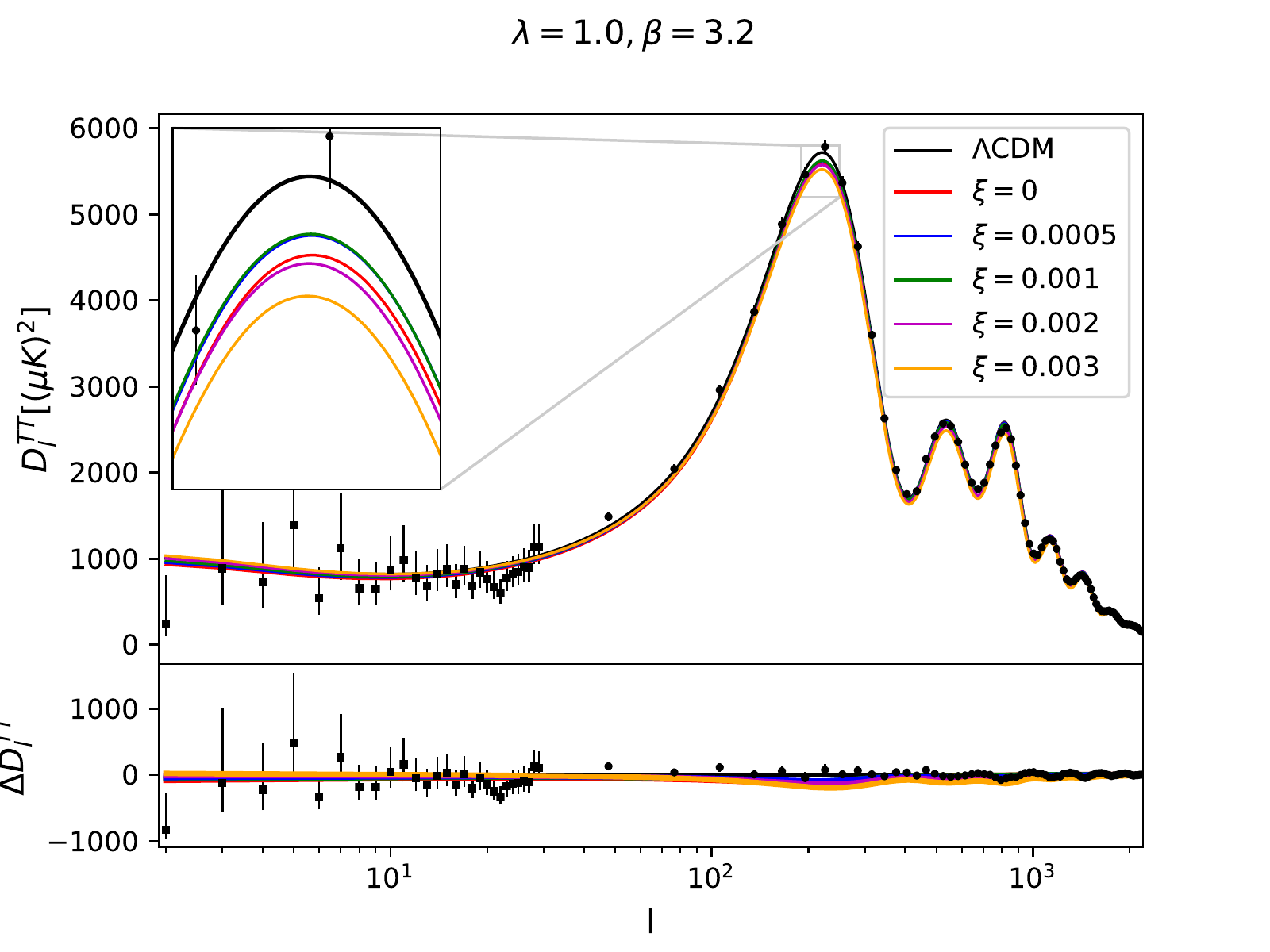}
	\caption{The temperature power spectrum for the non-minimally coupled $\beta$-inflation model. The different colors correspond to values of $\xi$ from zero to $0.003$, while we have considered the effect of both smaller and larger $\lambda$ ($\lambda=0.1$ - upper figures and $\lambda=1$ - lower figures), when $\beta=1.6$ or $\beta=3.2$. The black line represents the Planck (2018) $\Lambda$CDM model. }\label{fig4}
\end{figure*}

Fig.~\eqref{fig4} shows the predictions for the temperature power spectrum for different values of $\beta$, $\lambda$ and $\xi$. The predominant effect of these parameters is slightly changing the amplitude of the temperature power spectrum. We tested different combinations of $\beta$, $\lambda$, and $\xi$ and obtained the same minor variations in the amplitude. A non-minimal coupling enhances the effect of the shift in amplitude caused by the variation of $\beta$ and $\lambda$ when compared with the non-minimally coupled case, represented by the red line in all panels. Moreover, we can not discriminate among which parameter, $\beta$ or $\lambda$, is more critical to the shift in amplitude. Except that the effect of $\xi$ is more evident, we expect that data can well constrain the strength of the coupling of the field with gravity.

We perform the parameter estimation using the {\sc ModeCode} interfaced with the latest version of {\sc CosmoMC} code~\cite{Lewis:2002ah}, to check the viability of the model as a possible inflationary scenario.
The free parameters $\lambda$, $\beta$ and $\xi$ were assumed along with the usual cosmological parameters: $\left \{\Omega_bh^2~,~\Omega_ch^2~,~\theta~,~\tau\right \}$, i.e., the baryon and the cold dark matter density, the ratio between the sound horizon and the angular diameter distance at decoupling, and the optical depth, respectively. We also consider purely adiabatic initial conditions, fix the sum of neutrino masses to $0.06$ eV and the universe curvature to zero, and vary the nuisance foreground parameters~\cite{Planck:2019nip}.
Concerning the priors used for $\xi,\beta$, and $\lambda$, we have taken into consideration their impact on the predictions of the $n_s$ and $r$ parameters (see Fig. \ref{fig3}), and the discussion on the effect in the temperature power spectrum amplitude. The flat priors on the cosmological parameters used in our analysis are shown in Table~\ref{tab_priors}.

Regarding the observational data used in the analysis, we work with the CMB data from the latest Planck Collaboration (2018) release~\citep{Planck:2019nip}. Considering the high multipoles Planck temperature data from the 100-,143-, and 217-GHz half-mission T maps and the low multipoles data by the joint TT, EE, BB, and TE likelihood. EE and BB are the E- and B-mode CMB polarization power spectrum and TE is the cross-correlation temperature-polarization. Further, we use the tensor amplitude of B-mode polarization from 95, 150, and 220 GHz maps, coming from the Keck Array and BICEP2 Collaborations~\citep{PhysRevLett.114.101301,PhysRevLett.116.031302} analysis of the BICEP2/Keck field. Moreover, the combination with Planck high-frequency maps removes the polarized Galactic dust emission and constrains the parameters associated with the tensor spectrum (hereafter BK15).
Then, we combine the CMB data with Baryon Acoustic Oscillations coming from the 6dF Galaxy Survey (6dFGS)~\citep{bao1}, Sloan Digital Sky Survey (SDSS) DR7 Main Galaxy Sample galaxies~\citep{bao2}, BOSSgalaxy samples, LOWZ and CMASS~\citep{bao3}.

%
\begin{table}
\centering
\caption{Priors on the cosmological parameters considered in the analysis.}
{\begin{tabular}{|c|c|}
\hline
Parameter & Prior Ranges \\
\hline
$\Omega_{b}h^{2}$ & $[0.005 : 0.1]$ \\

$\Omega_{c}h^{2}$ & $[0.001 : 0.99]$ \\

$\theta$ & $[0.5 : 10.0]$ \\

$\tau$ & $[0.01 : 0.8]$ \\

$\xi$ & $[0 : 0.003]$ \\

$\beta$ & $[0.5 : 6]$ \\

$\lambda$ & $[0.1 : 2.0]$ \\
\hline
\end{tabular}\label{tab_priors}}
\end{table}

Finally, to perform a model comparison between the two cases of the $\beta$-exponential and the $\Lambda$CDM$+r$ models, we use the Deviance Information Criterion as statistical tool~\citep{Spiegelhalter:2002yvw}. This criterion investigates how well a model fits the data and if the available data favor its complexity. The DIC of a certain model is defined as
\begin{eqnarray}
DIC_\mathcal{M}\equiv -2\overline{\ln{\mathcal{L}(\theta)}} + p_D,
\label{27}
\end{eqnarray}
where the first term is the posterior mean of the likelihood $\mathcal{L}(\theta)$, namely the likelihood of the data given the model parameters, and the second term is the Bayesian complexity, $p_D=-2\overline{\ln{\mathcal{L}(\theta)}} + 2{\ln{\mathcal{L}(\bar{\theta)}}}$. The Bayesian complexity estimates the effective degrees of freedom in the model, which is quantified as the difference between the posterior mean likelihood and an estimator of model parameters maximizing the likelihood function, such that the model DIC is rewritten as $DIC_\mathcal{M}= -4\overline{\ln{\mathcal{L}(\theta)}} + 2{\ln{\mathcal{L}(\bar{\theta)}}}$. Hence, the DIC accounts for the goodness of fit and the Bayesian complexity of the model, such that models with more free parameters are penalized with a larger $p_D$.

The mean likelihood is a standard output from the chains of the MCMC analysis. Lastly, using the BOBYQA algorithm implemented in {\sc CosmoMC} for likelihood maximization, we can obtain the best-fit likelihood for each model.
Following \citep{Winkler:2019hkh} and previous works in the literature, the model selection is based on the lower DIC value being preferred concerning the reference value. This is quantified through the $\Delta DIC = DIC_{\mathcal{M}} - DIC_{ref} = 10/5/1$, meaning strong/moderate/null preference for the reference model, respectively.

\begin{table}
\centering
\caption{$68\%$ confidence limits and best-fit values for the cosmological parameters. The first columns-block show the constraints on the parameters of the $\Lambda$CDM+r model using the extended data set, i.e., the joint Planck 2018+BAO+BKP15 data. The second and third columns-block show the constraints on the parameters of the minimally coupled and non-minimally coupled models, using the extended data set Planck 2018+BAO+BKP15.
We indicate with $^{\ast}$ the derived parameters (except for the $n_s$ value of $\Lambda$CDM model, which is a primary parameter), and the last line shows the DIC values.}

\begin{tabular*}{\textwidth}{>{\footnotesize}c|>{\footnotesize}c|>{\footnotesize}c|>{\footnotesize}c|>{\footnotesize}c|>{\footnotesize}c|>{\footnotesize}c}
\hline
&\multicolumn{2}{c}{$\Lambda$CDM+r}
 &\multicolumn{2}{c}{Minimally coupled} &\multicolumn{2}{c}{Non-minimally coupled} \\
\hline
 {Parameter} & {mean} & {best fit} & {mean} & {best fit}& {mean} & {best fit}\\
\hline

{$\Omega_b h^2$}
& $0.02219 \pm 0.00019$
& $0.02243$
& $0.02230 \pm 0.00019$
& $0.02232$
& $0.02219 \pm  0.00018$
& $0.02216$	
\\
{$\Omega_{c} h^2$}
& $0.1192 \pm 0.0011$
& $0.1196$
& $0.1178 \pm 0.0009$
& $0.1177$
& $0.1193 \pm 0.0010$
& $0.1196$
\\
{$\theta$}
& $1.04096 \pm 0.00042$
&$1.04119$
& $1.04112 \pm 0.00040$
& $1.04119$
& $1.04096 \pm 0.00042$
& $1.04088$
\\
{$\tau$}
& $0.056 \pm 0.007$
& $0.057$
& $0.052 \pm 0.003$
& $0.054$
& $0.050 \pm 0.005$
& $0.047$
\\
{$\beta$}
& $-$
& $-$
& $1.85 \pm 0.95$
& $2.00$
& $2.73 \pm 1.44$
& $2.55$
\\
{$\lambda$}
& $-$
& $-$
& Unconstrained
& $-$
& Unconstrained
& $-$
\\
{$\xi$}
& $-$
& $-$
& $-$
& $-$
& $0.0012 \pm 0.0004$
& $0.0013$
\\
{$H_{0}^{\ast}$}
& $67.49 \pm 0.49$
& $67.65$
& $68.13 \pm 0.43$
& $68.21$
& $67.47 \pm 0.47$
& $67.29$
\\
{$\Omega_{m}^{\ast}$}
& $0.3120 \pm 0.0065$
& $0.3116$
& $0.3034 \pm 0.0056$
& $0.3024$
& $0.3122 \pm 0.0063$
& $0.3146$
\\
{$\Omega_{\Lambda}^{\ast}$}
& $0.6880 \pm 0.0065$
& $0.6884$
& $0.6966 \pm 0.0056$
& $0.6976$
& $0.6878 \pm 0.0063$
& $0.6854$
\\
{$n_s^{\ast}$}
& $0.9660 \pm 0.0039$
& $0.9664$
& $ 0.9860 \pm 0.0025$
& $0.9864$
& $ 0.9668 \pm 0.0042$
& $ 0.9656 $
\\
$r_{0.002}^{\ast}$
& $<0.024$
& $0.034$
& $0.045 \pm  0.019$
& $0.034$
& $0.017 \pm  0.011$
& $0.012$
\\
\hline
\hline
$\Delta$ DIC 
&\multicolumn{2}{>{\footnotesize}c}{Reference}
&\multicolumn{2}{>{\footnotesize}c}{5.8 }
&\multicolumn{2}{>{\footnotesize}c}{ -2.7}
\\
\hline
\end{tabular*} \label{tab:Tabel_results_1}
\end{table}

\section{Results}\label{sec5}

\begin{figure*}[t]
	\includegraphics[width=16cm]{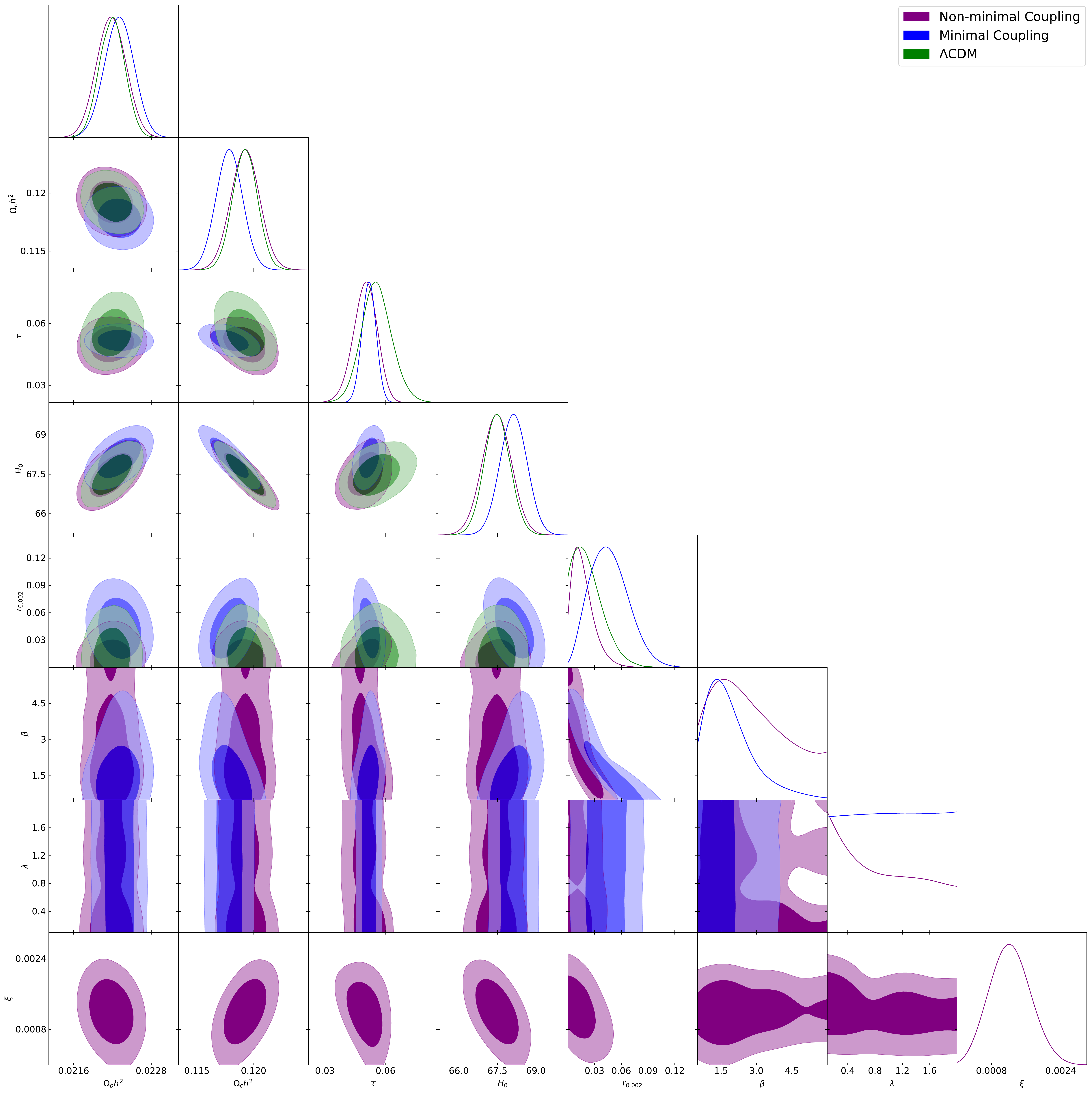}
	\caption{Confidence regions for the non-minimally and minimally coupled $\beta-$exponential models (purple and blue contours, respectively) and the reference $\Lambda$CDM model (green contours), all of them using the extended data set Planck 2018+BAO+BKP15.}\label{fig:triplot}
\end{figure*}

We summarize the main results for both minimally and non-minimally coupled $\beta-$ exponential models in Table~\eqref{tab:Tabel_results_1} and Fig.~\eqref{fig:triplot}. We show the constraints on the cosmological parameters and confidence {regions} for some parameters of interest. We updated the constraints on the minimally coupled model, considered previously in \citep{Santos:2017alg}, using the most recent CMB data combined with large-scale structure observations (Planck 2018+BAO+BKP15).

Considering both models investigated here, the constraints on standard cosmological parameters are in good agreement within $1\sigma$ with the Planck constraints~\citep{Akrami:2018odb}. The updated results on the minimally coupled model present $\beta=1.85\pm0.95$ and unconstrained $\lambda$. 
Noteworthy, $\lambda$ plays a minimal role in the amplitude of the temperature power spectrum, which is reflected in our updated results, and shows no degeneration with other parameters of the model. For the non-minimally coupled model, we have found $\beta=2.73\pm1.44$, unconstrained $\lambda$, and the coupling parameter  $\xi=0.0011\pm0.0004$, which shows a preference for the presence of a non-minimal coupling at the $1\sigma$ confidence level. Considering that we could not obtain a constraint on $\lambda$ in both models, we have tried to fix this parameter and investigate if the constraints on the other free parameters would improve. However, the analysis performed shows no improvement compared to that one with all the extra parameters kept free.

Despite the difference on $\lambda$ with the previous analysis for the minimally coupled model, the updated results we found continue to present a slightly smaller amount of total non-relativistic matter, resulting in a higher value of the dark energy density parameter $\Omega_\Lambda$ of $\Omega_\Lambda=0.6963\pm0.0055$. Regarding the inflationary parameters $n_s$ and $r$, we see that the best constraints on $r$ comes from the non-minimally coupled model, where $r=0.017\pm0.001$. Moreover, this value is within the $1\sigma$ limit from Planck, while the minimally coupled model predicts a higher value of $r=0.045\pm0.018$. Checking now the consistency with the $n_s$ limits, remember that the minimally coupled model has a bare agreement with the $2\sigma$ region, as seen in Fig.\eqref{fig2}, while the presence of a non-minimal coupling leads to lower $n_s$ at the same time that $r$ also decreases. We then obtain, using Eq.(\ref{24}), a value of $n_s=0.9668 \pm 0.0042$,  
\footnote{Note that for the value of $\lambda$ we use the arbitrary value 1 in the Eq.(\ref{24}), since this parameter is not constrained by the data.}
(second-to-last row in Table (\ref{tab:Tabel_results_1})) both in agreement within $1\sigma$ with the Planck constraints~\citep{Akrami:2018odb}.

In Fig.~\eqref{fig:triplot}, we show the confidence contours for all models analyzed. Note that we have a good agreement of both minimal and non-minimal models with the $\Lambda$CDM at $1\sigma$ C.L. (as also shown in Table \ref{tab:Tabel_results_1}). Finally, the model comparison performed using the DIC estimator is shown in the last line of Table \eqref{tab:Tabel_results_1}. According to the interpretative scale adopted, the minimally coupled model is strongly disfavored for the reference model. On the other hand, the non-minimal coupled scenario (7-parameters model) has a {moderate}  preference when compared with the $\Lambda$CDM model. The DIC value indicates how well a model fits the data and how much its complexity is adequate. Although the non-minimally coupled scenario is penalized due to its level of complexity, it is still capable of providing a good description of the data, as does the $\Lambda$CDM model.

\section{Conclusions}\label{sec6}

This paper introduces and investigates the observational viability of the non-minimally coupled $\beta-$exponential inflationary model. This class of models deserves to be analyzed as it is a generalization of the well-known power law inflation through the potential (\ref{12}), which is derived from the framework of braneworld scenarios~\cite{Santos:2017alg}. At the level of the $n_s-r$ plane, the minimally coupled model provides constraints that are in agreement with the confidence contours at $2\sigma$ level of the $\Lambda$CDM model, when using Planck 2015 data, but for the most recent {Planck 2018 + BAO measurements ($ns=0.9660 \pm 0.0039$ for  the $\Lambda$CDM+r model) this agreement with data essentially disappears ($ns=0.9860 \pm 0.0025$) 
as showed in Table \ref{tab:Tabel_results_1} and discussed in Sec. \ref{sec3}.} On the other hand, by introducing a non-minimal coupling of the field with gravity, we notice a clear improvement in the description of the data, as seen in Fig. \eqref{fig3}. This coupling is small, of order $\xi\sim 10^{-3}$, in agreement with other investigations, for different potentials \cite{Tenkanen2017}.

Regarding the statistical analysis, firstly, we updated the analysis presented in \cite{Santos:2017alg} for the minimally coupled $\beta-$exponential model with Planck 2018+BAO+BKP15 data, obtaining results on the standard cosmological parameters in agreement within $1\sigma$ C.L. with Planck results, a good constrain on $\beta$ but unconstrained $\lambda$. About the non-minimally coupled $\beta-$exponential scenario, we found a preference for the presence of a non-minimal coupling at the $1\sigma$ C.L., $\xi=0.0012\pm0.0004$, a good constrain on $\beta$ but also unconstrained $\lambda$. Finally, we performed a model selection considering the two classes of the $\beta-$exponential models and the $\Lambda$CDM scenario. In particular, we found that the minimally coupled $\beta$-exponential model is only marginally consistent with the data in $2\sigma$ C.L. if one considers only the $n_s-r$ plane (see Fig.\eqref{fig2}). Still, the DIC value strongly disfavored this scenario compared to the $\Lambda$CDM model. For the non-minimally coupled model, the inclusion of the $\xi$ parameter allows larger values of $\lambda$ bringing the predictions for the $n_s-r$ plane to an excellent agreement with the data. The improvement with the inclusion of non-minimal coupling, $\xi$, is confirmed by the statistical analysis, where we found a {moderately} preference for this scenario over the $\Lambda$CDM model.

Finally, it is worth mentioning that a study on the warm inflation scenario considering the $\beta$-exponential inflationary model is currently under investigation and will be reported in future communication.

\section*{Acknowledgements}

F.B.M. dos Santos is supported by Coordena\c{c}\~{a}o de Aperfei\c{c}oamento de Pessoal de N\'ivel Superior (CAPES). S. Santos da Costa acknowledges financial support from the Programa de Capacita\c{c}\~ao Institucional (PCI) do Observat\'orio Nacional/MCTI (grant no. 301869/2021-9).  R. Silva acknowledges financial support from CNPq (Grant No. 307620/2019-0). MB is supported by the Istituto Nazionale di Fisica Nucleare (INFN), sezione di Napoli, iniziativa specifica QGSKY. J. Alcaniz is supported by Conselho Nacional de Desenvolvimento Cient\'{\i}fico e Tecnol\'ogico CNPq (Grants no. 310790/2014-0 and 400471/2014-0) and Funda\c{c}\~ao de Amparo \`a Pesquisa do Estado do Rio de Janeiro FAPERJ (grant no. 233906). 
We also acknowledge the authors of the {\sc ModeCode} (M. Mortonson, H. Peiris and R. Easther) and {\sc CosmoMC} (A. Lewis) codes. This work was developed thanks to the High Performance Computing Center at the Universidade Federal do Rio Grande do Norte (NPAD/UFRN) and the Observat\'orio Nacional Data Center (DCON).

\bibliography{references}

\appendix

\end{document}